\documentclass[9pt,journal]{IEEEtran}
\usepackage{cite}
\usepackage{siunitx}

\ifCLASSINFOpdf
  \usepackage[pdftex]{graphicx}
\else
\fi

\usepackage[cmex10]{amsmath}

\usepackage{array}

\usepackage{url}

\begin{document}
\title{Suppressing Leakage Magnetic Field in Wireless Power Transfer using Halbach Array-Based Resonators}

\author{Yuichi~Honjo,
        Cedric~Caremel,~\IEEEmembership{Member,~IEEE,}
        Yoshihiro~Kawahara,~\IEEEmembership{Member,~IEEE,}
        and~Takuya~Sasatani,~\IEEEmembership{Member,~IEEE}

\thanks{This work was partially supported by JST ACT-X Grant Number JPMJAX190F, JSPS KAKENHI Grant Number 23H03378, and Value Exchange Engineering, a joint research project between Mercari, Inc. and the RIISE.}
\thanks{The authors are with the Graduate School of Engineering, The University of Tokyo, Tokyo, 113-8656, Japan (e-mail: honjo@akg.t.u-tokyo.ac.jp; sasatani@g.ecc.u-tokyo.ac.jp)}
\thanks{© 2023 IEEE. Personal use of this material is permitted.  Permission from IEEE must be obtained for all other uses, in any current or future media, including reprinting/republishing this material for advertising or promotional purposes, creating new collective works, for resale or redistribution to servers or lists, or reuse of any copyrighted component of this work in other works.}
\thanks{This is the accepted version of the manuscript. The final version is available at IEEE Antennas and Wireless Propagation Letters via the digital identifier DOI: 10.1109/LAWP.2023.3318244.}}

\markboth{IEEE ANTENNAS AND WIRELESS PROPAGATION LETTERS,~2023}%
{Shell \MakeLowercase{\textit{et al.}}: Bare Demo of IEEEtran.cls for IEEE Journals}

\maketitle

\begin{abstract}
Wireless power transfer has the potential to seamlessly power electronic systems, such as electric vehicles, industrial robots, and mobile devices.
However, the leakage magnetic field is a critical bottleneck that limits the transferable power level, and heavy ferromagnetic shields are needed for transferring large amounts of power.
In this paper, we propose a ferrite-less coil design that generates an asymmetric magnetic field pattern focused on one side of the resonator, which effectively reduces the leakage magnetic field.
The key to enabling the asymmetric field pattern is a coil winding strategy inspired by the Halbach array, a permanent magnet arrangement, which is then tailored for wireless power using an evolutionary strategy algorithm.
Numerical analyses and simulations demonstrated that the proposed coil structure delivers the same amount of power as spiral coils, while achieving an 86.6\% reduction in magnetic field intensity at a plane located 75~mm away from the resonator pair and a power efficiency of 96.0\%.
We verified our approach by measuring the power efficiency and magnetic field intensity of a test wireless power system operating at 6.78~MHz.
These findings indicate that our approach can efficiently deliver over 50~times more power without increasing magnetic field exposure, making it a promising solution for high-power wireless power transfer applications.
\end{abstract}


\section{Introduction}
\IEEEPARstart{W}{ireless} power transfer via magnetic field can seamlessly empower various electronics, but the leakage magnetic fields generated by these systems can cause negative effects, such as tissue heating and electromagnetic interference, which pose critical limitations on the transferable power level and application scenarios~\cite{effectleakagemagneticfield}.
Ferromagnetic shields have been used to suppress those fields~\cite{shieldbyferrite2}, but they are heavy and expensive and become lossy at frequencies exceeding a few MHz~\cite{hysteresis,metamate_shield}.
Therefore, there is a need for a resonator design that can suppress leakage magnetic fields without using ferromagnetic material and can also efficiently transfer power, especially for mobile applications where weight and cost are critical factors.

\begin{figure}[t]
    \centering
    \includegraphics[width=1\linewidth]{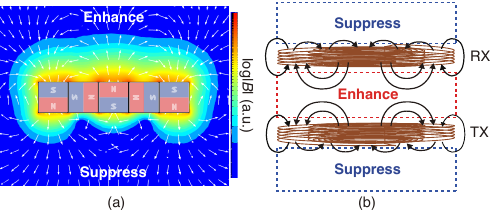}
    \caption{Suppressing the leakage magnetic field using a resonator structure inspired by the Halbach array.
    (a)~A Halbach array composed of five permanent magnets. The magnetic field on one side is enhanced, while it is suppressed on the other side.
    (b)~Suppressing the leakage magnetic field using resonators inspired by Halbach arrays.}
    \label{fig:halbach_intro}
\end{figure}

To address this need, we propose a ferrite-less resonator structure that generates an asymmetric magnetic field pattern focused on a single side of the resonator.
This design enables efficient wireless power transfer and leakage magnetic field suppression, with the key feature being a parametrically optimized coil pattern that vertically arranges numerous helix and spiral coils wound in both the clockwise and counterclockwise directions.
This winding generates a magnetic field pattern analogous to the Halbach array, an arrangement of permanent magnets that focuses the magnetic field to a single side of the array~\cite{Mallinson}~(see  Fig.~\ref{fig:halbach_intro}).
While prior literature has described coil structures~\cite{Halbachcoil} and a ferrite shield structure~\cite{Halbachshield} inspired by the Halbach array, the presented coil designs can not simultaneously achieve field suppression and high-efficiency power transfer because these simple imitations disregard critical factors, such as power losses and the difference in field distributions between coils and permanent magnets.
To overcome this limitation, we used the Covariance Matrix Adaptation Evolution Strategy (CMA-ES) algorithm to optimize the coil pattern for two design priorities: (a) leakage magnetic field suppression and (b) high power transfer efficiency.
We also demonstrated how varying the weights of these two design priorities can affect the performance of the resonators, and these results were verified through Method of Moments (MoM)-based simulations.
Finally, we built a wireless power transfer test system operating at 6.78 MHz and measured the leakage magnetic fields to confirm the results.

\section{Halbach array-based resonator}\label{sec:halbach-coil}
The Halbach array is an arrangement of permanent magnets, first proposed in 1973, which enhances the magnetic field of one side of the array, while suppressing the field on the other side~\cite{Mallinson,Halbach}~(See Fig.~\ref{fig:halbach_intro}(a)).
On the top side, the N and S poles of adjacent unit permanent magnets are clustered, virtually forming a large magnet that generates a far-reaching magnetic field.
Conversely, the N and S poles are isolated on the bottom side and behave as small magnets that generate a quickly decaying magnetic field.
Combined together, this makes the magnetic field focused on one side and suppressed on the other side of the magnet.

\begin{figure*}[t]
    \centering
    \includegraphics[width=1\linewidth]{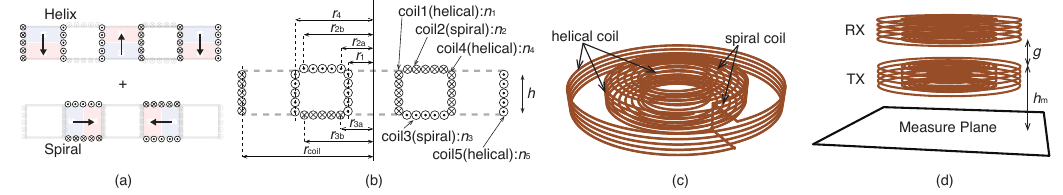}
    \caption{Structure of the Halbach array-based resonator.
    (a)~Replacing the permanent magnets of the Halbach array with helix and spiral coils.
    When the coils are combined, the magnetic field is enhanced on the upper side and suppressed on the lower side.
    (b)~Parameterization of the Halbach array-based resonator.
    $r_i$ are the radii of each coil, $n_i$ are the number of turns of each coil, and $h$ is the height of the resonator.
    (c)~The Halbach array-based resonator with all helix and spiral coils configured. This structure is a series connection of five helices and spiral coils with capacitors inserted in series.
    (d)~The position of the transmitter, receiver, and measurement plane, in which the magnetic field is measured.
    $g$ is the distance between the resonators, and $h_{\rm m}$ is the distance from the resonators to the measurement plane.}
    \label{fig:halbach_array}
\end{figure*}

For application to wireless power, the permanent magnets need to be replaced with inductors that form an oscillating magnetic field, and a field analogous to the Halbach array can be composed by combining helix and spiral coils with various polarities~\cite{Halbachcoil}.
We can observe from the cross-section of axis-symmetric coils that helix coils can imitate vertical magnets, and the spiral coils can imitate horizontal magnets~(See Fig.~\ref{fig:halbach_array}(a)).
By connecting these helix and spiral coils in the correct polarity, we can form a resonator that generates a field analogous to the Halbach array~(See Fig.~\ref{fig:halbach_array}(b)(c)(d)).

\section{Optimization method using an evolution strategy}\label{section-opt-process}
While it is possible to conceptually mimic the magnetic field distribution of a Halbach array using the coil structure outlined in section~\ref{sec:halbach-coil}, a straightforward design based on this concept may not be adequate to achieve high performance in wireless power systems.
Several critical factors for wireless power, such as power loss, are not considered, and the original Halbach array premises a 1-D array of square-shaped magnets rather than an axis-symmetric structure.
The field leakage at the array edges is significant, and the magnetic field distribution of coils and permanent magnets is different.
These issues are interdependent and challenging to address empirically.
Therefore, we proposed incorporating an evolution strategy algorithm in the design process.
A parameterized structure, an evaluation function, and an optimization algorithm are required to design a structure using an optimization algorithm.

Fig.~\ref{fig:halbach_array}(b) shows the parameterization of the coil structure. The coil comprises three helical coils and two spiral coils, encoded as a twelve-element vector $\Vec{a} = \{\Vec{r}, \Vec{n}, h\}$. This vector includes the radius vector $\Vec{r} = \{r_1,r_{2{\rm a}},r_{2{\rm b}},r_{3{\rm a}},r_{3{\rm b}},r_4\}$, the number of turns vector $\Vec{n}=\{n_1,n_2,n_3,n_4,n_5\}$, and the height $h$ (scalar). Note that the outer helical coil's radius is fixed to $r_{\rm coil}$, and the spiral coils (coils 2 and 3) have two radii, \textit{i.e.}, inner and outer radii. To ensure physical validity and prevent interference, we imposed the following restrictions on the geometry: the radii were bounded by $r_1<r_4<r_{\rm coil}$ and $r_{i{\rm a}} < r_{i{\rm b}} < r_{\rm coil}~(i = 2,3)$; the height was set within \SI{5}{mm}$<h<$\SI{100}{mm}; and the lower bounds of $r_1$, $r_{2{\rm a}}$, and $r_{3{\rm a}}$ were set to \SI{3}{mm}. The radius of the entire coil $r_{\rm coil}$ was fixed at \SI{150}{mm}, and the number of turns $n_i$ was restricted to integers between 0 and 5.

Secondly, the evaluation function was tailored to the specific requirements of the application, which for this study were (a) minimizing the leakage magnetic field and (b) maximizing power transmission efficiency.
To this end, we defined the evaluation function $E$ as a weighted sum of two factors: the maximum magnetic field intensity on a measuring plane $H_\mathrm{max}$ (as shown in Fig.\ref{fig:halbach_array} (d)) and the maximum power transfer efficiency $\eta_\mathrm{max}$.
The evaluation function can be expressed as:
\begin{align}
E = w H_\mathrm{max} - \eta_\mathrm{max}
\end{align}
Here, $E$ is to be minimized during the design process, and $w$ is a weighting factor determined based on the application's requirements.
In the optimization process, we calculated $H_\mathrm{max}$ and $\eta_\mathrm{max}$ using numerical analysis because performing electromagnetic field simulations is time-consuming and limits the feasible number of function evaluations.
We defined efficiency as the maximum efficiency $\eta_\mathrm{max}$ obtained when the load value is tuned to the value that maximizes power transfer efficiency, assuming the maximum efficiency point tracking technologies developed in previous studies\cite{efficiency_calculate}.
We calculated $\eta_\mathrm{max}$ based on the mutual inductance $M$ and the copper losses of the two resonators $r_1, r_2$, where we computed $M$ using Neumann's formula~\cite{neumann} and the copper losses by multiplying the wire length by the wire resistance of a unit length, considering the skin effect~\cite{calculate_method}.
$H_\mathrm{max}$ was the maximum magnetic field intensity in a measuring plane $h_m$ downward from the top of the transmitter, as shown in Fig.~\ref{fig:halbach_array}(d).
As this field intensity depends on the operating conditions, we calculated it for a power delivery of \SI{1}{W} to the load using Biot-Savart's law.

Finally, the optimization algorithm was selected based on the properties of the input parameters and the evaluation function. For this study, the input parameters were a mix of real numbers~(radii, height) and integers~(number of turns), and the computation of $H_\mathrm{max}$ required the selection of maximum value, which made the function non-differentiable.
Therefore, we used the Covariance Matrix Adaptation Evolution Strategy (CMA-ES)~\cite{CMAES,optuna,CMAESuse}, which is known to be efficient and converge quickly under such conditions.

\section{Optimization and Simulation results}

\begingroup
\renewcommand{\arraystretch}{1.5}
\begin{table}[t]
\caption{Parameters used for Optimization}
\label{table:setting}
\footnotesize
\centering
\begin{tabular}{wc{40mm}|wc{12mm}|wc{23mm}}
 Description & Symbol & Value \\
\hline
\hline
Operating frequency & $f_\mathrm{0}$ & 6.78~\si{MHz} \\
\hline
 Radius of the coil & $r_\mathrm{coil}$ & 150~\si{mm} \\
\hline
 Distance of measurement plane & $h_\mathrm{m}$ & 75~\si{mm}, 150~\si{mm} \\
 \hline
 Distance between the coils & $g$ & 20~\si{mm} \\
\end{tabular}
\end{table}
\normalsize
\endgroup

\begin{figure}[t]
    \centering
    \includegraphics[width=1\linewidth]{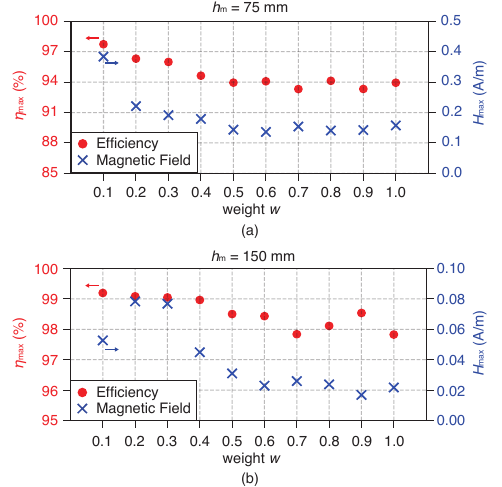}
    \caption{Results of the optimization process when the measurement plane position is defined as (a)~$h_{\rm m}=75\mbox{~mm}$ and (b)~$h_{\rm m}=150\mbox{~mm}$.
    The plot shows the performance of the optimized structure when the weight $w$ is varied from 0.1 to 1.0.
    The MoM-based simulation and experiment were conducted using a design resulting from a measurement plane of 75~\si{mm} and a weight $w$ of 0.3.}
    \label{fig:simulation_result}
\end{figure}

We conducted the optimization process by varying the weight $w$ and plotted the resulting power transfer efficiency and leakage magnetic field in Fig.~\ref{fig:simulation_result}.
To obtain these results, we iterated the optimization ten times with a limit of 10,000 function evaluations for each case and plotted the smallest value.
The parameters used for this evaluation are listed in Table~\ref{table:setting}.
Each optimization converged after approximately 2,000 function evaluations.
The figure demonstrates that a larger weight $w$ places more emphasis on reducing the leakage magnetic field, resulting in lower efficiency, and vice versa for a smaller $w$.

Next, we further investigated the optimized resonator structure and compared it with previous designs by conducting electromagnetic simulations using the method of moments~(MoM) simulations in FEKO.
As for the proposed design, we used the optimized design resulting from parameters $(w, h_\mathrm{m})=(0.3,75\mbox{~mm})$ and compared the leakage magnetic field and maximum power transfer efficiency with those of a normal spiral coil and a conventional Halbach array-based coil~\cite{Halbachcoil}.
Note that we used the distributed reactance compensation technique to compensate for the non-uniformity in the current that appears when operating coils with many turns at high frequencies~\cite{Distribute}.

\begin{figure}[t]
    \centering
    \includegraphics[width=1\linewidth]{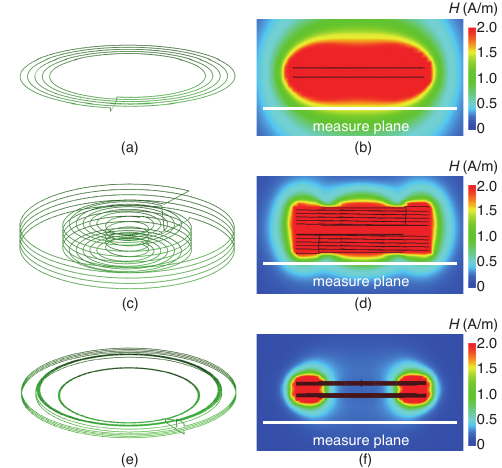}
    \caption{Resonator structure and the leakage magnetic field of (a)(b)~a 300~mm diameter, five-turn spiral coil, (c)(d)~the previous Halbach array-based coil~\cite{Halbachcoil}, and (e)(f)~the proposed Halbach array-based resonator.
    The power delivered to the load was the same~(1~W) for all configurations.
    }
    \label{fig:comparecoils_result}
\end{figure}

\begin{figure}[t]
    \centering
    \includegraphics[width=1\linewidth]{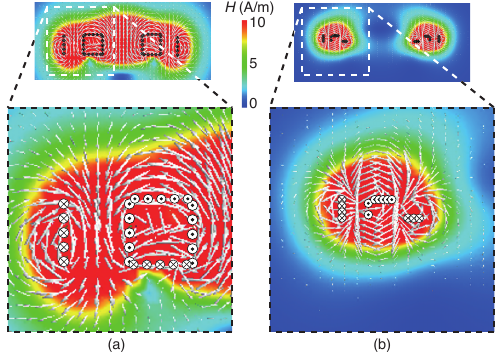}
    \caption{Comparison of the leakage magnetic field of Halbach array-based coils (a)~before and (b)~after optimization.
    A current of \SI{1}{A} was applied for both cases.
    The optimized resonator successfully suppressed the leakage magnetic field at the edges of the structure, whereas the non-optimized structure exhibited a far-reaching leakage magnetic field.
    }
    \label{fig:halbach_magneticfield}
\end{figure}

Fig.~\ref{fig:comparecoils_result} shows the leakage magnetic fields of the three compared coils when transferring \SI{1}{W} power to the load.
The figure demonstrates that the proposed design (Fig.~\ref{fig:comparecoils_result}(e)(f)) successfully suppressed the leakage magnetic field and outperformed the spiral coil (Fig.~\ref{fig:comparecoils_result}(a)(b)) and the conventional Halbach array-based coil (Fig.~\ref{fig:comparecoils_result}(c)(d)).
The maximum leakage magnetic field in the measurement plane $h_\mathrm{m} = \SI{75}{mm}$ was \SI{1.42}{A/m} for the spiral coil, \SI{0.97}{A/m} for the conventional Halbach array-based coil, and \SI{0.19}{A/m} for the optimized Halbach array-based coil. This means that the leakage magnetic field was suppressed by 87\% and 80\%, respectively, compared to the spiral coil and previous Halbach array-based coil.
By suppressing the leakage magnetic field, we could provide more current and deliver more power with the same leakage magnetic field. As delivered power is proportional to the square of the current and the current is proportional to the magnetic field, a magnetic field suppression of a factor of $\alpha$ suggests that $\alpha^2$ more power can be sent within the same leakage magnetic field. Therefore, these results indicate that the optimized Halbach array-based coil can transmit approximately 56 times more power than the spiral coil and 26 times more power than the conventional Halbach array-based resonator under the same leakage magnetic field.

The simulated power transfer efficiencies of the spiral coil, conventional Halbach array-based coil, and the proposed resonator were 99.3\%, 97.8\%, and 96.0\%, respectively.
The slight decrease in the efficiency of the proposed resonator was likely due to the increase in wire length (\textit{i.e.}, ESR) and the reduction of the magnetic fields in the power transfer direction as a side effect of significantly suppressing the leakage field.
Although the efficiency was slightly reduced compared to conventional designs, the significantly reduced magnetic field and the consequent increase in transferrable power levels make this a promising technology for high-power wireless power transfer applications.

Fig.~\ref{fig:halbach_magneticfield} presents a comparison of the leakage magnetic fields of a Halbach array before and after optimization.
As shown in Fig.~\ref{fig:halbach_magneticfield}(a), the original Halbach array exhibits a significant leakage near its edge, which causes to a far-reaching magnetic field below the array.
This is partly due to the lack of consideration of the array's edges in the general Halbach array design~\cite{Mallinson}.
Additionally, in the coil implementation, the magnetic field tends to concentrate near the wires, which can create larger leakage when the current paths are far apart, even though the field is intended to cancel out other magnetic fields.
In contrast, Fig.~\ref{fig:halbach_magneticfield}(b) demonstrates that the optimized Halbach array features current paths that are close together, while maintaining a field distribution similar to that of the original design.
This qualitative observation indicates that the interdependent challenges pointed out in section~\ref{section-opt-process} were successfully resolved by incorporating an evolution strategy algorithm in the design. These factors allowed the previously mentioned increase in transferable power levels of more than 50 times.

\begin{figure}[t]
    \centering
    \includegraphics[width=1\linewidth]{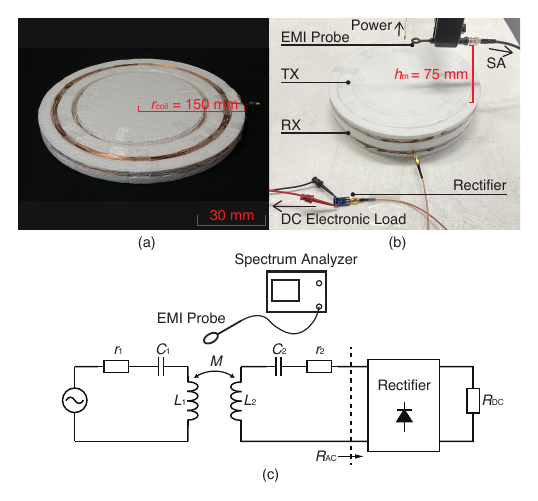}
    \caption{
    (a)~The fabricated resonator structure, (b)~measurement setup for magnetic field measurement~(SA: spectrum analyzer), and (c)~circuit diagram of the experiment setup. We used an electronic load to tune $R_{DC}$ to the optimum load presented in Table \ref{table:value}. Note that we modeled the distributed reactance compensation~\cite{Distribute} as a single equivalent capacitor for simplicity. 
    }
    \label{fig:measurement_result}
\end{figure}

\begingroup
\renewcommand{\arraystretch}{1.5}
\begin{table}[t]
\caption{Circuit parameter values of Fig. \ref{fig:measurement}}
\label{table:value}
\footnotesize
\centering
\begin{tabular}{wc{30mm}|wc{10mm}|wc{10mm}}
 Description & Symbol & Value \\
\hline
 ESR of TX & $r_\mathrm{1}$ & 4.07~\si{\Omega} \\
 ESR of RX & $r_\mathrm{2}$ & 3.78~\si{\Omega} \\
 Capacitance of TX & $C_\mathrm{1}$ & 24.6~\si{\pico\farad} \\
 Capacitance of RX & $C_\mathrm{2}$ & 23.8~\si{\pico\farad} \\
 Inductance of TX & $L_\mathrm{1}$ & 22.4~\si{\micro H} \\
 Inductance of RX & $L_\mathrm{2}$ & 23.2~\si{\micro H} \\
Mutual inductance & $M$ & 1.65~\si{\micro H} \\
Optimum load & $R_\mathrm{DC}$ & 83.2~\si{\Omega} \\
\end{tabular}
\end{table}
\normalsize
\endgroup

\section{Measurement}
\begin{figure}[t]
    \centering
    \includegraphics[width=1\linewidth]{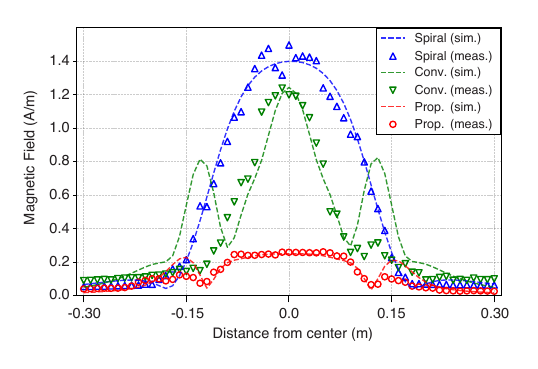}
    \caption{The measured and simulated magnetic fields of a spiral coil, the conventional Halbach array-based coil, and the proposed resonator. We measured the $z$-axis direction magnetic field at extracted points. The measured results showed that the experiment and simulation corresponded well and that the proposed design can significantly suppress the leakage magnetic field. (Conv.: conventional Halbach coil~\cite{Halbachcoil}; Prop.: proposed Halbach coil; sim.: simulation; meas.: measurement.)
    }
    \label{fig:measurement}
\end{figure}

This paper next proceeds to describe the process of building a wireless power system to test the optimized resonator's design and validate the numerical analysis and electromagnetic simulations.
To fabricate the resonators, we wound \SI{1}{mm} diameter copper wire around a styrofoam core~(Fig.~\ref{fig:measurement_result}(a)).
The measured power transmission efficiencies of the spiral coil, the conventional Halbach array-based coil, and the proposed resonator were 95.9\%, 97.2\%, 93.0\%, respectively, which suggests that the proposed structure can achieve a transfer efficiency comparable to the previous approaches. These efficiency values were slightly lower than the simulated results due to the losses introduced by the non-ideality of the fabrication process.
We then measured the leakage magnetic field when transferring power. Fig.~\ref{fig:measurement_result}(a) shows the fabricated coil, and Fig.~\ref{fig:measurement_result}(b) shows the measurement setup.
The experiment's diagram is shown in Fig.~\ref{fig:measurement_result}(c), and Table~\ref{table:value} shows the circuit parameters derived via fitting the measured impedance matrix.
A near-field EMI probe connected to a spectrum analyzer was used to measure the leakage magnetic fields.
We used a DC electronic load $R_\mathrm{DC}$ to tune the full-bridge rectifier's input impedance close to the optimum load value $R_\mathrm{AC}$ used in the design process and to measure the power delivered to the load.
In the measurement, we adjusted the transmitter power to deliver 1~W of power to the rectifier, assuming an 89\% AC-DC conversion efficiency of the full-bridge rectifier based on circuit simulation results using LTSpice.
Fig.~\ref{fig:measurement} shows the measured and simulated leakage magnetic fields of a spiral coil, the conventional Halbach array-based coil, and the proposed resonator in the measurement plane $h_m = \SI{75}{mm}$. Note that the magnetic field in the direction perpendicular to the resonator was compared for simplicity. These measurement results highlight that the experiment and simulation corresponded well and that the proposed design can significantly suppress the leakage magnetic field.

\section{Conclusion}
This paper presented a ferrite-less resonator structure inspired by the Halbach array that enables efficient wireless power transfer and leakage magnetic field suppression.
Simulation and measurements showed that the CMA-ES-based optimization process could address the non-ideal factors that emerge when leveraging Halbach array-based resonators in wireless power transfer.
These findings indicate that our approach can efficiently deliver over 50~times more power without increasing magnetic field exposure.
One advantage of this approach is that it could be used together with conventional shielding techniques to achieve equivalent or even stronger shielding with less shielding material by mitigating the requirement for magnetic field attenuation. Thus, investigating the combination with ferromagnetic material to achieve lightweight and secure shielding is a promising direction of future work.


\begin{thebibliography}{1}

\bibitem{effectleakagemagneticfield}
H.~Kim, J.~Cho, S.~Ahn, J.~Kim and J.~Kim, ``Suppression of leakage magnetic field from a wireless power transfer system using ferromagnetic material and metallic shielding,'' \emph{2012 IEEE International Symposium on Electromagnetic Compatibility}, pp. 640-645, 2012.

\bibitem{shieldbyferrite2}
W.~Zhang, Q.~Yang, Y.~Li, Z.~Lin, M.~Yang and M.~Mi, ``Comprehensive analysis of nanocrystalline ribbon cores in high-power-density wireless power transfer pads for electric vehicles,'' \emph{IEEE Transactions on Magnetics}, vol. 58, no. 2, pp. 1-5, 2022.

\bibitem{hysteresis}
W. Roshen, ``Ferrite core loss for power magnetic components design,'' \emph{IEEE Transactions on Magnetics}, vol. 27, no. 6, pp. 4407-4415, Nov. 1991.

\bibitem{metamate_shield}
J.~Besnoff, M.~Chabalko and D.~S.~Ricketts, ``A frequency-selective zero-permeability metamaterial shield for reduction of near-field electromagnetic energy,'' \emph{IEEE Antennas and Wireless Propagation Letters}, vol. 15, pp. 654-657, 2016.

\bibitem{Mallinson}
J.~Mallinson, ``One-sided fluxes – a magnetic curiosity?'' \emph{IEEE Transactions on Magnetics}, vol. 9, no. 4, pp. 678-682, 1973.

\bibitem{Halbachcoil}
H.~Kim, K.~Hwang, J.~Park, D.~Kim and S.~Ahn, ``Design of single-sided AC magnetic field generating coil for wireless power transfer,'' \emph{IEEE Wireless Power Transfer Conference (WPTC)}, pp. 1-3, 2017.

\bibitem{Halbachshield}
X.~Chen, Z.~Guo, J.~Jiang, H.~Jiang and H.~Chen, ``Ultra-broadband near-field magnetic shielding realized by the Halbach-like structure,'' \emph{Applied Physics Letters}, 120.19, 2022.

\bibitem{Halbach}
K.~Halbach, ``Design of permanent multipole magnets with oriented rare earth cobalt material,''
\emph{Nuclear Instruments and Methods}, vol. 169, no. 1, pp. 1-10, 1980.

\bibitem{efficiency_calculate}
M.~Kato, T.~Imura and Y.~Hori, ``New characteristics analysis considering transmission distance and load variation in wireless power transfer via magnetic resonant coupling,'' in \emph{2012 IEEE International Telecommunications Energy Conference (INTELEC)}, pp. 1-5, 2012.

\bibitem{neumann}
T.~Imura and Y.~Hori, ``Maximizing air gap and efficiency of magnetic resonant coupling for wireless power transfer using equivalent circuit and neumann formula,'' \emph{IEEE Transactions on Industrial Electronics}, vol. 58, no. 10, pp. 4746-4752, 2011.

\bibitem{calculate_method}
T.~Sasatani, Y.~Narusue and Y.~Kawahara, ``Genetic algorithm-based receiving resonator array design for wireless power transfer,'' \emph{IEEE Access}, vol. 8, pp. 222385-222396, 2020.

\bibitem{CMAES}
N.~Hansen, S.~D.~Müller and P.~Koumoutsakos, ``Reducing the time complexity of the derandomized
evolution strategy with covariance matrix adaptation (CMA-ES),'' \emph{Evolutionary Computation},
vol. 11, no. 1, pp. 1-18, 2003.

\bibitem{optuna}
T.~Akiba, S.~Sano, T.~Yanase, T.~Ohta and M.~Koyama, ``Optuna: a next-generation hyperparameter optimization framework,'' \emph{Proc. 25th ACM SIGKDD International Conference on Knowledge Discovery \& Data Mining (ACM KDD '19)}, pp. 2623-2631, 2019.

\bibitem{CMAESuse}
M.~D.~Gregory, Z.~Bayraktar and D.~H.~Werner, ``Fast optimization of electromagnetic design problems using the covariance matrix adaptation evolutionary strategy,'' \emph{IEEE Transactions on Antennas and Propagation}, vol. 59, no. 4, pp. 1275-1285, April 2011.

\bibitem{Distribute} 
Y.~Narusue and Y.~Kawahara, ``Distributed reactance compensation for printed spiral coils in wireless power transfer,'' \emph{2017 IEEE Wireless Power Transfer Conference (WPTC)}, pp. 1-4, 2017.

\end{thebibliography}
\end{document}